# Self-Diagnosis through Chatbot-based Symptom Checkers: User Experiences and Design Considerations

Yue You, MS[1], Xinning Gui, PhD[1]
[1]Pennsylvania State University, State College, PA, USA

**Abstract**

*Recently, there has been a growing interest in developing chatbot-based symptom checker (CSC) apps in the healthcare market. CSC apps provide potential diagnoses for users and assist them with self-triaging based on Artificial Intelligence (AI) techniques using human-like conversations. Despite the popularity of such CSC apps, little research has been done to investigate their functionalities and user experiences. To do so, we conducted a feature review, a user review analysis, and an interview study. We found that the existing CSC apps lack the functions to support the whole diagnostic process of an offline medical visit. We also found that users perceive the current CSC apps to lack support for a comprehensive medical history, flexible symptom input, comprehensible questions, and diverse diseases and user groups. Based on these results, we derived implications for the future features and conversational design of CSC apps.*

**Introduction**

Recently CSC (chatbot-based symptom checker) apps have proliferated in the mobile application market. A chatbot is defined as a computer program that can conduct a conversation with humans. A CSC app assesses medical symptoms using chatbots that have conversations with users[1]. Users input their symptoms and acquire diagnoses by communicating with the chatbot embedded in the CSC app. Most CSC apps have appeared in the last few years following the development of the AI algorithms that were applied to conversational chatbots[2]. Based on our calculations, some CSC apps (e.g., Ada, K Health, and HealthTap) have been downloaded from the Google Play store more than one million times since the release. Despite their popularity, however, CSC apps, as a consumer-facing diagnostic technology, have received little attention in the healthcare domain; most previous research has focused on diagnostic tools for clinical decision support, helping medical professionals or healthcare providers seek information[3] and reduce diagnostic errors[4].

Although developers of CSC apps promise various benefits, such as support for medication information[5] and triage decisions[6], users, especially individuals with high-risk diseases, can put their lives at risk if they blindly trust the apps' diagnostic results[7]. CSC apps can lead to inaccurate predictions and unintended consequences in medical diagnoses[8]. Thus, users' perceptions of the CSC apps' effectiveness are essential as the apps can influence users' further healthcare decisions. Plenty of studies have focused on reviewing health applications, such as exploring the applications' socio-technical implications[9], suggesting guidelines for CSC apps' implementations[10], and giving an overview of the chatbot-based health apps used for behavioral changes[9]. Other studies have explored the conversational design of healthcare chatbots, illustrating the importance of considering users' design preferences[10]. However, few studies have focused primarily on the features and the effectiveness of CSC apps. Little research has also been done to recommend guidelines for healthcare chatbots' conversational design from the users' perspectives. As it is still unclear what features CSC apps should provide to effectively assess users' symptoms, there is a crucial need for an empirical understanding of how users perceive the effectiveness of CSC applications.

To fill these gaps, we combined a feature analysis, an app review analysis, and semi-structured interviews to look at users' needs and perceptions of CSC apps in the hopes of influencing future design. We conducted a feature review on eleven CSC apps to analyze their functionalities and analyzed a set of online reviews taken from the U.S. Apple and Google Play stores to examine users' experiences. We then conducted ten interviews with CSC app users to cross-validate the online review analysis results. We have identified five CSC app deficiencies: an insufficient consideration of health history, rigid input requirements, problematic probing questions, the ignorance of diverse health conditions, and a lack of functions regarding follow-up treatments. To the best of our knowledge, this is the first study to review the functionalities and users' experiences of CSC apps. Our findings reveal new challenges for the functional and conversational design of CSC apps from users' perspectives.

## Methods

We began our work with a feature analysis on selected CSC apps. Based on the feature analysis, we used a review analysis to examine user perceptions for the identified features. We then conducted ten semi-structured interviews to cross-validate our review analysis findings.

*App selection*

To obtain qualified CSC apps for the feature review, we searched the Apple app and Google Play stores for CSC apps (free, freemium, and paid) that are accessible within the U.S., using the following keywords: "medical OR health OR healthcare AND chatbot OR symptom checker." The initial search returned 70 apps on the Apple app store and 289 on Google Play. We then selected 11 apps for the feature analysis on the basis of the following six criteria: (1) the presence of a chatbot function; (2) the use of English; (3) medical diagnosis as the main focus; (4) a general health focus (not only mental health); (5) health consumers and not medical professionals as the target users; (6) downloadable and functional. We excluded 71 apps that did not employ a chatbot, another 200 that specifically targeted mental health, and 77 that did not focus on symptom checking (Figure 1). The 11 apps we selected are Ada, K Health, Ask NHS, Your.MD, Mediktor, HealthTap, Apothēka Patient, Sensely, Health Buddy, Babylon, and NHS online: 111. All but "Health Buddy" can be found in both the Apple app and Google Play stores; "Health Buddy" can only be found in the Apple app store.

From the 11 CSC apps used for the feature review, we selected 4 apps (Ada, K Health, Your.MD, and Ask NHS) for the review analysis. Our selection was based on the two following criteria: (1) the apps must have at least 700 reviews; (2) the apps must have at least 1000 ratings. This was done to eliminate any potential bias of a particular user.

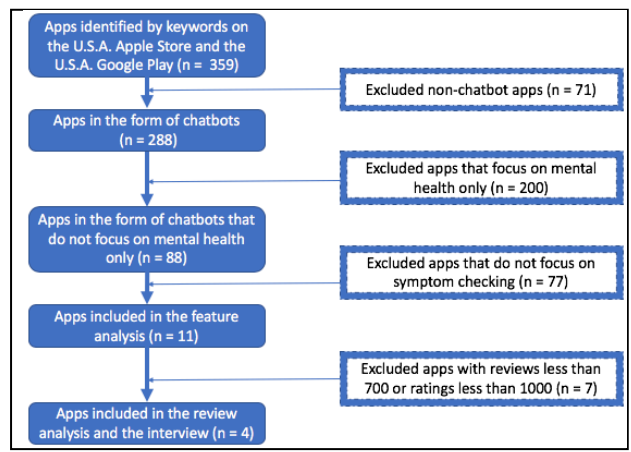

**Figure 1.** App selection flow graph

*App feature analysis*

A feature review is a common research approach in health informatics that involves reviewing app history (e.g., ratings and developers)[11,12], data policies[13], functions[14], and target users[15] to evaluate the validity of mobile apps. Following the app feature analysis methodology, we carried out a critical feature analysis to shed light on the feature design of the CSC apps. We downloaded all 11 apps, studied their main features, and then coded them based on the eight stages of the offline diagnostic process[16]: (1) establishing a patient history, (2) conducting a physical exam, (3) evaluating symptoms, (4) giving an initial diagnosis, (5) ordering further diagnostic tests, (6) performing and analyzing test results, (7) providing a final diagnosis, and (8) providing referrals or other follow-up treatments. Finally, we developed a coding sheet of feature categories with subcategories.

*App review analysis and semi-structured interview analysis*

We first conducted a review analysis for the four selected CSC apps, Ada, K Health, Your.MD, and Ask NHS. A review analysis is a process widely used in health informatics to identify how users rated an app[11,17]. App reviews exist in publicly available datasets and cover a large user population. The app reviews can reveal users' experiences with and attitudes toward apps[18]. We followed the app review analysis methodology, downloading 500 most-recent reviews for each CSC app with the help of two app analytics tools (ASO100 and App Annie) and analyzed these reviews critically. To cross-validate the online review analysis results and acquire more in-depth insights as to how users perceive CSC apps, we conducted ten semi-structured interviews with CSC app users. Participants were recruited through social media and *Studyfinder,* a web-based participant recruitment tool provided by Penn State Clinical Translational Science Institute. We first utilized a screening survey to filter out unqualified participants. The eligibility criteria for users were as follows: (1) over the age of 18; (2) has used CSC app(s) to seek possible diagnoses; and (3) the last time of use was within the last year. We conducted interviews until the data from the review analysis and the interview study reached "theoretical saturation[19]." We finally recruited ten interviewees (see Table 1). Then we conducted semi-structured individual interview sessions with each participant. These interviews took about 30 minutes to 1 hour, starting with the following open-ended questions: (1) which app do you prefer (if the user has used more than one)? (2) how do you feel about having conversations with these apps? (3) what are your preferred forms of conversation? Participants could decline to answer any question, refuse to be audio-recorded, or withdraw from the study at any time. We then manually transcribed the interviews.

We then used thematic analysis in an inductive approach[20] for the online reviews and interview transcripts. We first familiarized ourselves with the reviews and then generated the initial codes. The initial code list was comprised of over twenty codes. We then searched for themes and acquired a collection of candidate themes and sub-themes. Next, we reviewed and refined our themes to ensure internal homogeneity and external heterogeneity[21]. Finally, we acquired our final thematic map and defined five themes: 1.) how users perceived the profile function of these CSC apps; 2.) how users considered the input function; 3.) how users interpreted the probing questions; 4.) how users understood the diagnostic results; and 5.) what functions users desired regarding follow-up treatments. When reporting our findings, we used R1, R2, etc. to indicate the different users writing online reviews in our review analysis and P1, P2, etc. to denote each interview participant.

**Table 1.** Participants' demographic information.

| #   | Age | Gender | CSC Apps                       | Recruiting Methods |
|-----|-----|--------|--------------------------------|--------------------|
| P1  | 23  | Female | Ada, K Health, Your.MD, Ask NHS | Social media      |
| P2  | 29  | Female | Ada, K Health, Your.MD, Ask NHS | Social media      |
| P3  | 24  | Female | K Health, Ask NHS              | StudyFinder        |
| P4  | 24  | Male   | Ada, K Health, Ask NHS         | Social media       |
| P5  | 28  | Female | Ada, K Health, Ask NHS         | Social media       |
| P6  | 27  | Female | Ada, K Health                  | Social media       |
| P7  | 26  | Female | K Health                       | Social media       |
| P8  | 27  | Male   | K Health                       | Social media       |
| P9  | 25  | Female | Ada, K Health                  | Social media       |
| P10 | 40  | Female | Your.MD                        | Social media       |

**Results**

*App feature analysis*

After coding the features mapped by the eight stages of the offline diagnostic process, we categorized their features (Table 2). We found most apps can support the following five processes: establishing a patient history, evaluating symptoms, giving an initial diagnosis, ordering further diagnostic tests, and providing referrals or other follow-up treatments. These apps do not support conducting physical exams, providing a final diagnosis, and performing and analyzing test results because these three processes are difficult to realize using mobile apps.

To establish patient history, six apps vary in the amount of personal, demographic, and health information users need to input. For example, HealthTap requires exhaustive amounts of information, including name, age, gender, ethnicity, current and past conditions, medications, allergies, vaccinations, lab test results, treatment, lifestyle, and pregnancy status. On the contrary, other apps require only basic information. An example is Your.MD, which only requires name, date of birth, gender, and consultation history in its profile.

To give an initial diagnosis, assessment presentations vary. Three apps (Ada, K Health, and HealthTap) reflect the likelihood of diagnostic results in comparison with similar users; four apps (Ada, K Health, Your.MD, and Mediktor) employ pictures or diagrams to explain conditions; only one app (Ask NHS) uses videos to explain diagnostic results.

For referrals or other follow-up treatments, the apps usually connect with offline medical services, including contacting doctors (K Health, Mediktor, HealthTap, and Apothēka Patient), getting prescriptions (Your.MD and K Health), and finding pharmacies (Ask NHS and Your.MD) and hospitals (NHS online: 111). Symptom tracking is also prevalent in these apps (Ada, K Health, and Your.MD); users can utilize this function to track the severity of their symptoms over time.

**Table 2.** Features of the eleven apps.

| Name | 1.Establish A Patient History | 3.Evaluate Symptoms | 4.Give an Initial Diagnosis | 5.Order Further Diagnostic Tests | 8.Referrals or Other Follow-up Treatments |
|---|---|---|---|---|---|
| Ada | Date of birth, gender, height, weight, medication, allergies, health background (e.g., diabetes) | √ | Likelihood, cause-and-effect diagram | × | Symptom tracking |
| K Health | Date of birth, gender, height, weight, medications, allergies, ethnicity, smoking, surgeries, chronic conditions, family history | √ | Likelihood, pictures for conditions | × | Chat with doctors, get prescriptions, symptom tracking |
| Ask NHS | × | √ | Text / video | × | Find pharmacies |
| Your.MD | Date of birth, gender, consultations history | √ | Text / pictures for conditions | Order tests using third-party apps | Get prescriptions, find pharmacies, symptom tracking |
| Mediktor | Age, gender, height, weight, medication, race, allergies, risk factors, past medical history, past surgical history | √ | Text / pictures for conditions | × | Chat with doctors |
| HealthTap | Age, gender, medications, allergies, ethnicity, current and past conditions, vaccinations, lab test results, treatment, lifestyle, pregnancy | √ | Likelihood, text | × | Chat with doctors |
| Apothēka Patient | Age, gender, weight, height, blood group | √ | Text | × | Book an appointment with a doctor |
| Sensely | × | √ | Text | × | × |
| Health Buddy | × | × | Text | × | × |
| Babylon | × | √ | Text | × | × |
| NHS online: 111 | × | √ | Text | × | Find hospitals |

In addition to different functionalities, these apps employ diverse forms of information expression. Most apps allow users to use text to input their symptoms. However, the flexibility of input is limited in certain apps (Ada, K Health, Ask NHS, Your.MD, Mediktor, and Sensely), which ask users to choose from a restrictive list. Only two apps (Ask NHS and Sensely) allow users to input information via voice recording. Two apps (K Health and HealthTap) let users input information by clicking on separate parts of an image. Most apps use various forms of text, voice, and pictures to explain the questions presented during conversation. K Health uses cartoons and Ada uses pictures of real human body parts to explain the medical jargon that appears in the questions. For example, K Health uses a cartoon picture to illustrate the red joint and Ada uses a picture of a real human throat to elucidate a reddened throat (Figure 2). Some CSC apps also use humanizing language in their questions, such as "*Thanks for telling me about your…*" (K Health). Two apps (Ask NHS and Sensely) employ avatars with a recorded human voice to ask questions (Figure 2).

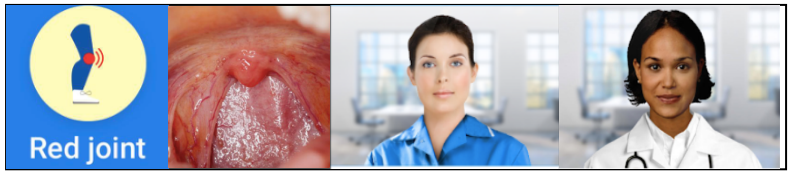

**Figure 2.** Example screenshots illustrating the pictures and avatars embedded in the CSC apps (from left to right: a cartoon picture of a red joint from K Health, a picture of a reddened throat from Ada, the avatar of Ask NHS, the avatar of Sensely).

*App review analysis and interview analysis*

After the feature analysis, we analyzed users' reviews of four CSC apps–Ada, K Health, Ask NHS, and Your.MD. We also conducted semi-structured interviews to study users' perceptions. Our analysis revealed users' experience in the following aspects: a failure to consider patient history, rigid input requirements, problematic probing questions, a neglect of diverse health conditions and user groups, and a lack of enough functions for follow-up treatments. We summarize the findings in Table 3 and report details in this section.

**Table 3.** Summary of main findings from online reviews and interview transcripts.

| Perspectives | Details | Diagnostic processes |
|---|---|---|
| A failure to consider Patient history | Ignore existing diagnoses and medications | 1.Establish a patient history |
|  | Overlook personal information |  |
| Rigid input requirements | Limited descriptions for disease dimensions | 3.Evaluate symptoms |
|  | Limited presentations of symptoms input |  |
| Problematic probing questions | The difficulty of comprehending complex sentences | 3.Evaluate symptoms |
|  | Slow interactions with numerous or seemly irrelevant questions |  |
|  | Unreasonable asking sequence |  |
| A neglect of diverse health conditions and user groups | Lack support for diverse diseases | 4.Give an initial diagnosis |
|  | Lack support for certain user groups |  |
| A lack of enough functions for follow-up treatments | Need to support customized symptoms tracking | 8.Referrals or other follow-up treatments |
|  | Need to reach out medical professionals |  |
|  | Need to provide information regarding medications |  |

(1) A failure to consider Patient history

Although in our feature analysis, we found that some CSC apps could record the basic health information of users, some perceived that none of the records were sufficient when compared with those required by offline medical consultations. Some users noted that many CSC apps did not ask for existing diagnoses and current medications, which they deemed essential information. For instance, R1 had a desire to list existing diseases in the profile. P3 thought that medications and surgeries were important to reach a diagnosis, especially for people who had relevant conditions.

*"The only thing I would suggest is to have a way to include existing diagnosed conditions in your profile such as anxiety or asthma etc..."* (R1; Ada)

*"if someone is like actually having a lot of medication and surgeries, the profile is important. Because they're actually having surgeries and medications that might be relevant to their current symptoms."* (P3; K Health)

Users also found that some CSC apps overlooked some personal information, such as race. R2 thought that race should be considered when making a diagnosis. Sickle cell disease (SCD) is a genetic disease[22], common among people whose ancestors are from sub-Saharan Africa, Spanish-speaking regions in the Western Hemisphere, and Mediterranean countries. Thus, this disease is strongly connected to racial information, which made R2 skeptical of the accuracy of K Health because K Health did not ask his/her race.

*"Why wouldn't a health diagnosis app ask one's race? I was curious to see if you could diagnose my known disease, Sickle Cell Anemia. Yet without asking my race, it was almost impossible to do so…"* (R2; K health)

(2) Rigid input requirements

Some users also found that some CSC apps could not process a complexity of symptoms, especially when dealing with diseases with a large array of symptoms or symptoms defined by users. In the feature analysis, we discovered that most CSC apps ask users to choose symptoms from a preset list. Users thought that these lists restricted their input.

First, some complained about the limited descriptions for disease dimensions. In the following example, R3 desired to input further details related to symptom frequency. R3 wanted additional options, such as "sometimes," to the current choice of "yes" or "no". R3 explained, *"... but there has to be an option for "at times" as a response to a symptom question instead of a "yes, no or idk" response…"* (R3; Ada).

Moreover, some users wanted to input multiple locations for their symptoms. As R4 posted, *"…Could do with a multiple-choice response on some answers, my pain is not located to just 1 area…"* (R4; Ask NHS). R5 had a similar take: *"The area of symptoms also only generally asks chest, legs, hands. There are other parts of the body also…"* (R5; Ada). In these two instances, the users thought that detailed location information was essential to reach a more accurate diagnostic result.

Second, some users were dissatisfied with the limited input for symptom presentation, especially in apps that had a pre-structured symptom selection list (e.g., K Health). P1 complained that K Health couldn't recognize the symptoms she input. She recounted, *"when I typed in 'hyperthyroidism', it didn't respond. I realized you have to input common symptoms, like 'tired'. I also typed in 'heart rate', but it didn't have this word. I know they can't recognize all words; the thing is I don't know what they have or not have in their database"* (P1; K Health). She believed that K Health was unable to identify particular medical terms, such as hyperthyroidism and heart rate; however, she was unsure what words she needed to use to describe her symptoms. Other users also found that some CSC apps lacked support for certain medical terms. In R6's case, Your.MD wouldn't acknowledge the simple medical term, "pimple." R6 explained, *"... it just does not understand what a simple 'pimple' means, suggests me skin rash and many more irrelevant information over and over again…"* (R6; Your.MD). R6 wanted to search for information about a "pimple"; however, Your.MD did not understand it as a medical term and gave the wrong answers.

(3) Problematic probing questions

Even though our feature analysis found that several CSC apps provide explanations for medical jargon, some users still had trouble understanding the probing questions and complained about a large number of them.

First, our interviewees mentioned the difficulty of comprehending the complex language the apps frequently use. For example, P2 thought that the questions Ask NHS asked were difficult to understand. She said, *"The questions of ASK NHS are not that readable, like 'How does general behavior or thoughts seem at the moment?' So just use some simple words to ask"* (P2; Ask NHS). She thought that Ask NHS should use simpler terminology so that users could better understand the questions.

Second, some users complained about the quantity of questions, especially with Ada and K Health, which made interactions with the CSC apps very slow and lengthy. R7 complained that Ada and Ask NHS had too many questions to answer, with no diagnostic results following the questionnaire. R7 wrote, *"[I]…was not given a diagnosis at the end of her hundred questions..."* (R7; Ask NHS).

P5 also disliked answering so many questions. For example, K Health asked her whether she had allergies and high blood pressure. P5 thought that these questions made the conversation with the app much longer than a consultation with an offline doctor. She noted, *"I think they are asking me too many questions, they are trying to cover every single possible aspect. The app was checking whether I have any allergy because I was coughing. And also like that high blood pressure or something. But the real doctors did not involve these symptoms because they can just look at me"* (P5; K Health). Additionally, P5 doubted the necessity of some of the questions. She said, *"So say I'm coughing, it's asking me 'do you have pain in the muscle?' I was [also] wondering why I need to tell them about blood pressure if I'm just coughing, sometimes I wonder why they ask these questions?"* (P5; K Health). She felt confused about why K Health asked these questions regarding the pain and blood pressure and she did know the relation between these questions and her symptom (i.e., cough). In line with their opinions, P10 perceived the long-winded questions bothered her because of their unreasonable asking order. She stated, *"It should ask highly relevant first and then ask lowly relevant questions. But Your.MD, its questions are not logical. When I say I'm vomiting, it asked you first, like 'have you had more alcohol then you think your body can cope with?' Then it asked me 'do you have any of these symptoms today?' Then it asked, 'Have you drunk any alcohol today?' So, I think these questions are not logical. The questions are too many and inaccuracy"* (P10; Your.MD).

Adding to the difficulty was that many users found the chatbots' conversational style problematic. R8 thought that Ada's questioning was rigid and lacked humanity: *"Conversation stilted and unlike a real human at all"* (R8; K health).

Some wished that their interactions with the CSCs more closely mimicked real conversations. For example, R9 wanted Ask NHS to express emotion and have an increased response flexibility: *"...need more interaction and give one the ability to express their feelings more than prepared answers…"* (R9; Ask NHS).

(4) A neglect of diverse health conditions and user groups

While all CSC apps in the feature analysis could provide initial diagnoses, some users in the review analysis noted CSC apps lacked the sufficient knowledge to support diverse diseases and special user groups.

First, some users perceived CSC apps should have supported certain particular diseases. For example, R10 intended to acquire information regarding skin burns and throbs from Ask NHS. However, R10 simply received answers regarding menstruation, which were off track from his/her intended search. As R10 noted, "*Absolutely useless app I wanted answers to why my skin burns and throbs for a long period of time her answers were about a f*****g period?!?!*" (R10; Ask NHS). This implies Ask NHS cannot recognize certain medical terms and has difficulty in understanding users' input from the user' perspective.

Specifically, some CSC apps showed support for chronic diseases, yet more comprehensive knowledge of the complexity of chronic conditions was desired by some users. For example, some users perceived these AISC apps could not distinguish chronic diseases from acute diseases efficiently. The below review shows R11 perceived Ask NHS could not identify the difference between chronic kidney disease and acute kidney disease.

"*There are major deficiencies in this app. For example, it is hazy about the terms chronic and acute. I asked about chronic kidney disease. And it responded about acute kidney infection, an entirely different condition. I shall remove the app as soon as I finish this comment*." (R11; Ask NHS)

R12 noted that the app lacked the knowledge base to provide information on his/her chronic diseases, OAB (overactive bladder) and IC (interstitial cystitis).

"*…this app doesn't know what OAB is nor IC which are two chronic illnesses I have …*" (R12; K Health).

Second, some users also noticed that the existing design of the apps did not accommodate diverse user groups. R13 wanted Your.MD to recognize the experiences particular to the mothers of babies. R13 wrote, *"...this app doesn't take into consideration women with young babies when asking about sleep patterns*" (R13; Your.MD).

Those identifying as transgender showed their dissatisfaction with the CSC apps. R14 complained that Ada overlooked the health issues of trans people and did not offer more gender setting options.

"*…the poor handling of terminology that is needlessly binary and unaccommodating of trans health issues (like increased incidence of breast cancer after transitioning as an AMAB trans woman) are actively HARMFUL to trans people - not to mention the unpleasant and dysphoria-inducing experience that I MUST input my "sex at birth" with no other options, though they may be medically relevant...*" (R14; Ada).

In addition, some online users wanted to use the CSC apps for children. R15 thought Ask NHS ignored situations where caregivers would use it with their kids.

"*Wanted to use it for my children but there is no option for that. Can only use it for one person and not children.*" (R15; Ask NHS)

(5) A lack of enough functions for follow-up treatments

Some users expressed their approval of the existing functions regarding the follow-up treatments that we identified in the feature analysis. For example, R16 thought that the symptom tracking function was useful to track headaches.

"*Amazing to be able to track my symptoms for a couple of weeks and finally understand what I can do to get better. Now no more headaches*." (R16; Your.MD)

R17 was satisfied to acquire other medical services from Ask NHS. He/she stated, "*Useful. People using this need to understand you are not getting a diagnosis, it can help to guide you to the right place or person to speak to about your problem or concern. Self-care option very good.*" (R17; Ask NHS)

However, some users also showed their desire for more follow-up treatment functions, such as tracking customized symptoms. This was something R18 complained about. "*Can't track customized symptoms*" (R18; Your.MD).

Some desired a medication information function. R19 stated, "*Only downside is that it doesn't offer specific medicines*" (R19; Ada). For apps that do not provide this function, a follow up with medical professionals was desired by some users, as shown in this review, "*Please add the option to contact or connect to doctors and professionals and provide them with your past reports and assessments to track your progress*" (R20; Ada).

**Discussions**

We found that most CSC apps can support five diagnostic processes that are included in the common offline medical visit: establishing a patient history, evaluating symptoms, giving an initial diagnosis, ordering further diagnostic tests, and providing referrals or other follow-up treatments. Through our review analysis and interview study, we discovered that users found weaknesses in the CSC apps' following functions: health history establishment, symptoms input, probing questions presentation, and diverse disease and user group support. Users also wished for more comprehensive features, such as medication information and customized symptom tracking. We will discuss these findings in this section.

*Design app features to bridge the social-technical gap.* Our study identifies that there is a social-technical gap between the features of CSC apps and offline medical consultations during each diagnostic process that the CSC apps support. When establishing a patient history, a clinical probing process usually collects a complete medical history and reviews a patient's previous activities[16]. However, the profiles of the CSC apps missed critical information in their medical history, which disappointed users. As such, the current profile of CSC apps needs to be more comprehensive. We also found the CSC apps' knowledge base ill-suited to disease complexity during the evaluation of symptoms. For example, when inputting symptoms, users found it difficult to input the sufficient dimensions of their conditions. However, in the offline setting, describing symptoms is complicated, requiring details regarding frequency, severity level, and location. Additionally, the real-world probing process requires communication skills, including empathy and the building of rapport[23]. Users complained about the stilted language of the apps' probing questions. Although some CSC apps employ humanizing language, work is still needed to more closely mimic human conversation. Finally, users perceived that the CSC apps lacked support for diverse users, echoing the guidelines of inclusive design, which requires product design to take into account the needs of all users without requirements for adaption[24]. The inclusiveness of offline healthcare services has already been emphasized[25]. Our study reveals that we should also consider inclusiveness in CSC apps design, reporting that users wanted CSC apps to accommodate diverse user groups.

These findings offered a new angle to consider in future consumer-facing diagnostic tools design and research: users' experiences can be improved if these tools provide similar functions and experiences to clinical visits in the real world. Future design should strive to take into consideration users' offline experiences, enhance the technology's knowledge base, and follow the guidelines of inclusive design to accommodate diverse contexts. In the case of the CSC apps, establishing a comprehensive health profile, allowing users to input different levels of symptoms, improving response speed, and providing more functions for follow-up treatments could improve the user experience.

*Address users' needs in the conversational design of healthcare chatbots.* Our findings reported that users had concerns about the input limitations and the comprehensibility of language in the conversational design of the CSC apps.

First, our study reported that users had difficulties in inputting symptoms. Users found that some CSC apps could not recognize their input, such as medical terms. Previous research has found that the difficulty of describing symptoms accurately can affect the users' acceptance of chatbots[26] and that accurately recognizing human input plays an important role in chatbot design[27]. Our study confirms these arguments as our participants desired flexible input in the chatbot conversations. Users wanted to input symptoms more easily and to be able to use words that they perceived as simple and familiar. This calls for research on approximate string matching and character recognition in the conversational design of healthcare chatbots.

Second, our findings reported that users perceived some language difficult to understand. Our study found that users had difficulty understanding the medical jargon that appeared in questions, which may lead to inappropriate input from users. Previous studies have emphasized the importance of decreasing the language complexity used in healthcare systems, such as the Electronic Health Records (EHR)[28] and the medication management systems of older patients[29] used in offline medical visits. Few studies have examined the users' perceptions of the language complexity of online healthcare chatbots. Our study, however, reveals that health chatbots also need to increase the comprehensibility of their probing questions based on users' perceptions, as a great number of people do not have a sufficient health literacy to interact with these healthcare systems[28]. In addition, users felt confused about the sequence of questions and the relationship among these questions; they did not know how these questions were related to the diagnostic results. Thus, information regarding why CSC apps ask certain questions, how these questions are related, and how these questions relate to the diagnostic results should be explained to users. This indicates that we should provide an explanation of the CSC apps' AI model to users. Previous studies drew on the social sciences and HCI knowledge to explain AI models[30,31] to users. While these studies provide insight as to what can constitute a good explanation, limited attention has been paid to users' needs for the explanations themselves. Since users are the ones to evaluate explanations, in contrast to the studies that propose researcher-driven explanations, our study highlights the users' requirements for explanations in conversation with healthcare chatbots.

Drawing from our findings and discussions, the future conversational design of healthcare chatbots should consider how to improve the input flexibility and the presentation of probing questions. First, the design should improve the functions of approximate string matching and character recognition and assist users to input their symptoms. Second, healthcare chatbots should use comprehensible language and provide explanations during conversations.

*Implications for clinical practice*. Our study indicates that CSC apps are prevalent in digital online app stores. Since users can self-diagnose using these apps, health risks may arise if users blindly trust their diagnostic results. Health providers may guide users by introducing the credibility as well as the limitations of these apps to assist users to make more reliable decisions.

**Conclusions**

Our study aims to explore the effectiveness of CSC apps and understand users' perceptions of them. By reviewing the functions of CSC apps, studying user reviews, and conducting user interviews, we have identified the features that consumer-facing diagnostic tools should provide for users and have shed light on the future conversational design of healthcare chatbots. However, our study also has some limitations. First, the Apple app and the Google Play stores' reviews do not include users' demographic information or distribution, so we were unable to factor this into our analysis. Second, not all users of the CSC apps posted a review online, resulting in a potential selection bias. Third, we only conducted ten interviews, which is a small sample. Although the combined interview and app review data did reach the theoretical saturation, larger scale studies need to be conducted in the future to acquire more comprehensive information regarding users' perceptions of CSC apps and their effectiveness.